\documentclass[
    ,final            
  ]
  {aipproc}

\layoutstyle{8x11single}

\newcommand{\osz}{\ensuremath{^1\textrm{S}_0}}

\newcommand{\tpz}{\ensuremath{^3\textrm{P}_0}}

\newcommand{\tso}{\ensuremath{^3\textrm{S}_1}}

\begin{document}

\title{Nucleon-Nucleon Interactions from the Quark Model}


\author{C.~Downum\footnote{Presenting Author}~}{
  address={Clarendon Laboratory, University of Oxford, Parks Road, Oxford, 
OX1 3PU, UK}
}

\author{J.~R.~Stone}{
  address={Clarendon Laboratory, University of Oxford, Parks Road, Oxford, 
  OX1 3PU, UK}
  ,altaddress={Department of Physics and Astronomy, University of Tennessee, 
  Knoxville, TN 37996, USA}
}

\author{T.~Barnes}{
  address={Department of Physics and Astronomy, University of Tennessee, 
  Knoxville, TN 37996, USA}
  ,altaddress={Physics Division, Oak Ridge National Laboratory, Oak Ridge, 
  TN 37831, USA} 
}
\author{E.~S.~Swanson}{
  address={Department of Physics and Astronomy, University of Pittsburgh, 
  Pittsburgh, PA 15260, USA}
}
\author{I.~Vida\~na}{
  address={Centro de F\'{i}sica Computacional. Departamento de F\'{i}sica.
Universidade de Coimbra, Rua Larga, 3004-516 Coimbra, Portugal}
}

\begin{abstract}
We report on investigations of the applicability of non-relativistic
constituent quark models to the low-energy nucleon-nucleon (NN)
interaction.   The major innovations of a resulting NN potential are
the use of the $^3$P$_0$ decay model and quark model wave functions to
derive nucleon-nucleon-meson form-factors, and the use of a colored
spin-spin contact hyperfine interaction to model the repulsive core
rather than the phenomenological treatment common in other NN potentials. 
We present the results of the model for
experimental free NN scattering phase shifts, S-wave
scattering lengths and effective ranges and deuteron properties.  
Plans for future study are discussed.
\end{abstract}

\maketitle


\section{Introduction}

The NN interaction has been the object of study for over 50 years.
The development of the highly accurate potentials in 1990s showed that the NN interaction can be quantitatively expressed very precisely although the physical mechanism of the interaction has yet to be understood.
Most of these models use a partial-wave decomposition of the NN potential and vary potential forms and parameterizations in dependence on partial wave channel.
The NN potentials are either thought to rise from one-boson exchange (OBE) (e.g. Nijmegen and Bonn families) or are constructed of different functional forms multiplied by general operators (central, tensor, spin-orbit etc) which are then parameterized (Argonne and Urbana families), with one-pion exchange included to model the long-range part of the potential.  
Chiral perturbation theory is a modern alternative approach which attempts, to derive the NN-force from contact interactions and pion exchange as the low-energy QCD degrees of freedom.
The Nijm I\cite{stoks1994chq}, Argonne v18\cite{wiringa1995ann}, N$^3$LO\cite{entem2003acd}, and CD-Bonn\cite{machleidt2000hpc}~potentials are used in this work for comparison.  

In this work, we present a potential motivated from the Non-Relativistic Quark Model, the Oxford potential.
It is shown to be able to well reproduce NN scattering phase shifts data and the properties of the deuteron.

This work is organized as follows.  
First, we briefly review two quark models of strong interaction dynamics.
Secondly, we create the Oxford potential using these models.
Then the results of our parameter tuning are presented for NN data and we show the high momentum behavior of the potential.
Finally, we give our conclusions and plans for future work.

\section{Quark Models of Strong Interactions}

\subsection{$^3$P$_0$ Model}
\begin{figure}
  \includegraphics[width=4in]{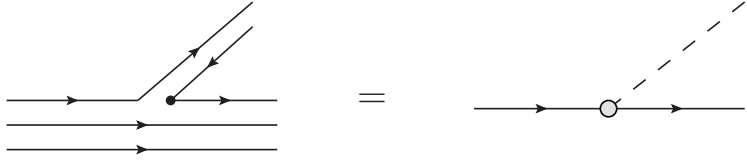}
  \caption{\label{fig:3P0Eff} A diagram illustrating the technique for calculating quark model form-factors for use in meson exchange potentials.}
\end{figure}
It was first suggested by Micu\cite{micu1969drm} that hadron decays might be dominated by the production of a $q\bar q$ pair from the vacuum with $J^{PC}=0^{++}$ quantum numbers, i.e. they are in a $^{2S+1}L_J =$ ~\tpz~state.  
Her idea has been applied to successfully estimate 3-point vertices with one additional meson in the final state\cite{leyaouanc1973npq}.  

Fig. \ref{fig:3P0Eff}~illustrates how to parameterize the effective field theory from the microscopic decay diagram using the~\tpz~model.  
External hadron simple harmonic oscillator SHO wave functions are attached to the microscopic decay diagrams to account for non-perturbative hadronization effects and yields the following expressions for the form-factor $\mathcal{F}$: 
\begin{equation}
  \mathcal{F}(\vec{p_i},\vec{p_f}) = \exp\left\{ - \frac{\left( \vec{p_f} + \vec{p_i} \right)^2}{24\left( 3\beta^2 + \alpha^2 \right)} - \frac{\left( \vec{p_f} - \vec{p_i} \right)^2}{6\alpha^2} \right\}.
\end{equation}
$\alpha, \beta$ are the SHO wavefunction widths for the nucleons and meson respectively and $p_i$, $p_f$ are the initial and final state momenta.

\subsection{One Gluon Exchange}

\begin{figure}
  \includegraphics[width=2in]{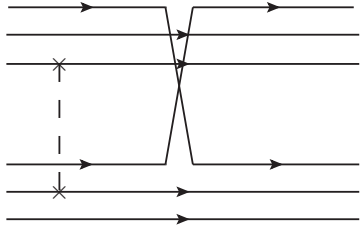}
  \caption{\label{fig:OGE} A representative diagram for the quark model OGE potential.  The $\times$ denote the colored spin-spin contact interactions.}
\end{figure}

Attempts to model the NN potential via a OGE interaction between bound state hadrons dates back to Liberman in 1977\cite{liberman1977srp}.
The potential due to the OGE appears to be dominated by the contact spin-spin hyperfine interaction (for recent reviews see Shimizu {\it et al.}\cite{shimizu2000snn} and  Valcarce {\it et al.}\cite{valcarce2005qms}).

Barnes {\it et al.} used the ``quark Born diagram'' formalism to model the dominant contact spin-spin interaction with constitute interchange (OGE+CI) to preserve color neutrality. 
A typical diagram for the process is present in Fig. \ref{fig:OGE}.
The OGE+CI interaction is an alternative model of short range NN dynamics -- compared with the usual heavy meson exchange model.
Also, it illustrates that the different scales of physics can be self-consistently incorporated into a potential using the quark model.

\section{The Oxford Potential}
For the Oxford potential, we modeled the NN interaction as a combination of one pion exchange (OPE), one sigma exchange (OSE) and the OGE+CI potential.
For the OPE and OSE, we use the standard the one meson exchange mechanism,
with the form-factors calculated from the \tpz~model. 
Charge dependence was incorporated as in CD-Bonn\cite{machleidt2000hpc}.

In order to fit the P and F wave phase shifts, additional repulsion was needed in these channels.  
Therefore, we added one omega exchange in these channels with a pure vector, isovector coupling in these channels.
A detailed presentation of the potential will be the subject of a forthcoming paper\cite{downum2010qml} 
and the quark level origins of the $\omega$ exchange effects will be investigated.

\section{Results}
The parameters of the Oxford potential were fitted to the single-energy values of the NN scattering phaseshifts as determined from the SAID analysis program\cite{SAID} and to the properties of the deuteron. 
No attempt has been made to perform a fit to NN scattering data of this 'proof of principle' potential. 
As the potential does not include Coulomb and higher-order electromagnetic corrections, only data in the $np$ scattering channel have been considered in the fit.

\subsection{Scattering Phase Shifts}
\begin{figure}[tp]
  \includegraphics[angle=0,width=5in]{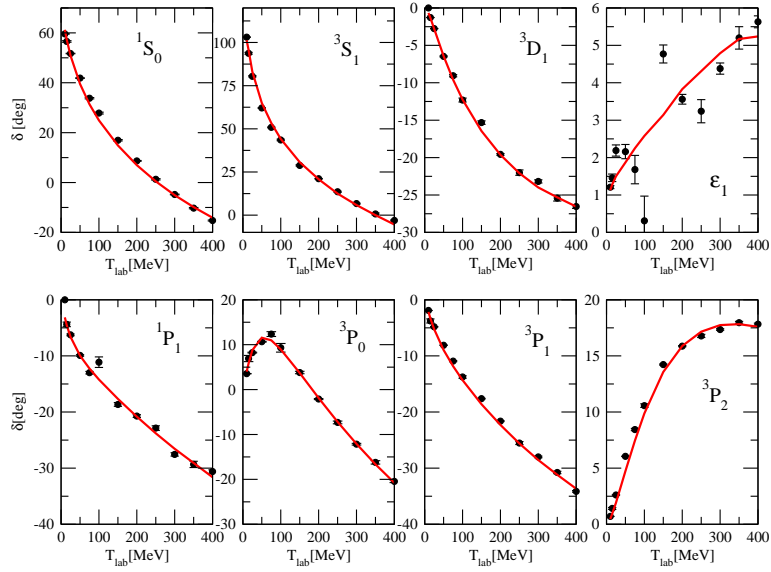}
  \caption{\label{fig:PSTable} Phase shift and inelasticity predictions of the Oxford model for the S, P, and $^3$D$_1$ partial waves and the $\epsilon_1$ mixing angle.  Data from the SAID program\cite{SAID} are the black dots while the predictions are the solid red curves.}
\end{figure}

The NN potential is non-perturbatively strong. 
Therefore, we solve a partial wave decomposed Lippmann-Schwinger type equation.  
The phase shifts are extracted from the resulting $T$ matrix using standard techniques.  
We present the results of our phase shift parameterization in Fig. \ref{fig:PSTable}.
The results of the highly accurate potentials are not presented -- all the potentials reproduce the experimental phase shifts well.

As can be seen from Fig. \ref{fig:PSTable}, the Oxford potential is able to reproduce the low-lying phase shifts very well.  
We only considered data up to $T_{\rm Lab} = 400$MeV.
Higher $L$ phase shifts are not presented here, but achieve a comparable fit.

\subsection{Deuteron Properties}
The deuteron is the only NN bound state.
We solve the Schr\"odinger Equation to calculate the bound state properties of the deuteron.
The deuteron properties of the Oxford potential are presented along with the results of the other highly accurate potentials and the experimental values.
Detailed discussion of the experimental values may be found elsewhere\cite{machleidt2000hpc}.
The mass of the sigma was fine-tuned to reproduce the binding energy, the other properties are as followed without subsequent adjustment.

Most quantities are accurate to within 1\% when the experimental error is considered.  
Given the simplicity of the model and its constraints, we find these results highly encouraging.
In particular we draw attention to the quadrupole moment which the Oxford potential is able better reproduce than the highly accurate potentials.

\begin{table}[pth]
  \centering
 \begin{tabular}{ccccccc}
    Property & Oxford & CD-Bonn & Nijmegen I & Argonne v18 & N$^3$LO & Experiment\\ \hline
    $B_d$[MeV]                        & 2.2246 & 2.224575  & 2.224575 &  2.224575 & 2.224575 &  2.224575$\pm$0.000009\\
    $A_S$[fm$^{1/2}$]                 & 0.8918   & 0.8846    & 0.8841  &  .8850  & 0.8843 & 0.8848$\pm$0.0009\\
    $\eta$                            &0.0262  & 0.0256    & 0.0253  & .0256   & 0.0256 & 0.0256$\pm$0.004\\
    $\sqrt{\langle r^2 \rangle}$[fm]  &1.9767 & 1.966     & 1.9666  &  1.967  & 1.978 & 1.971$\pm$0.006\\
    $Q_d$[fm$^2$]                     &0.2871  & 0.270     & 0.2719  &  .270   & 0.275 & 0.28590$\pm$0.00030\\
    $P_D$ [\%]                        & 5.604  & 4.85      & 5.664    &   5.76 & 4.51  & 
  \end{tabular}
  \caption{\label{tab:Deuteron} The deuteron properties of the Oxford, CD-Bonn, Nijmegen I, 
  and Argonne v18 potentials along with experimental values.  Relativistic and Meson Current corrections
  have not been included.}
\end{table}

\subsection{S-wave Scattering Lengths and Effective Ranges}
Our results for scattering lengths, $a^N$, and effective ranges, $r^N$, due to the strong nuclear force for the S-wave channels are summarized in Table~\ref{tab:SLER} in comparison to experiment and predictions of the other highly accurate potentials.
The \osz~channels required the introduction of small charge symmetry breaking in addition to charge independence breaking.
In the \tso~channel, the values of the $a^N_t$ and $r^N_t$ quoted were obtained using parameters fitted only to the binding energy of the deuteron.
We find the discrepancy of the order of 1\%~between predictions of the Oxford potential and experiment encouraging.

\begin{table}[pth]
\begin{tabular}{ccccccc}
         & {   Oxford   } & {   CD-Bonn   } & {   Nijmegen I   } & {   Argonne v18   } & N$^3$LO &  {   Experiment   }\\ \hline
	        \multicolumn{6}{c}{$^1$S$_0$}  \\
$a^N_{pp}$ &        -17.365 & -17.4602    &              &  -17.164    & -17.083 &                      \\
$r^N_{pp}$ &          2.886 &   2.845     &              &    2.865    &  2.876  &                      \\
$a^N_{nn}$ &        -18.949 & -18.9680    &              &  -18.818    & -18.900 & -18.9   $\pm$0.4     \\
$r^N_{nn}$ &        2.859   &   2.819     &              &    2.834    & 2.838   &   2.75  $\pm$0.11    \\
$a^N_{np}$ &        -23.85  & -23.7380    &              &  -23.084    & -23.732 & -23.74  $\pm$0.020   \\
$r^N_{np}$ &        2.753   &   2.671     &              &    2.703    & 2.752   &   2.77  $\pm$0.05    \\
\multicolumn{6}{c}{$^3$S$_1$} \\
$a^N_t$      &        5.46  &   5.4196   &     5.4194   &     5.402   & 5.417 & 5.419  $\pm$0.007   \\
$r^N_t$      &        1.77  &   1.751    &     1.7536    &     1.752  & 1.752 &   1.753  $\pm$0.008
\end{tabular}
\caption{\label{tab:SLER} The Scattering Lengths and Effective Ranges for the Oxford, CD-Bonn, 
Nijmegen I\cite{deswart1995len}, 
Argonne v18, and N$^3$LO potentials along with experimental values.}
\end{table}

\section{High Momentum Behavior}
It has been known for some time that the highly accurate potentials remain strong even for high values of momentum inducing strong correlations in many-body wave functions.
Sophisticated techniques have been developed to integrate out the unphysical high momentum degrees of freedom while retaining the physical, low momentum behavior of the potentials\cite{coraggio2008smc,bogner2006srg}.

\begin{figure}[tp]
  \includegraphics[angle=0,width=3in]{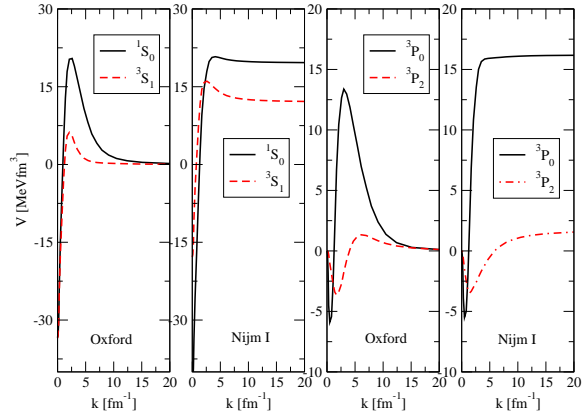}
  \caption{\label{fig:MomBehavior} The Oxford and Nijm I on-shell potential matrix elements for a range of momentum.} 
\end{figure}

In Fig. \ref{fig:MomBehavior}~we plot the potential matrix elements in the S and triplet P channels as functions of momentum for the Oxford and Nijm I potentials.  
Because of the quark model wave functions the Oxford potential has a non-local Gaussian form-factor which strongly suppresses the potential as momentum increases.
This form-factor is similar to the one employed in the chiral N$^3$LO potential.
The rapid fall off in our potential, as opposed to say Nijm I, is physically preferred.

\section{Conclusions}
We have been conducting research into the possibility of calculating NN interactions mechanisms from the quark model.
To demonstrate that this technique can lead to an accurate, physically motivated potential, we have produced the Oxford potential.

In this paper we reported on the preliminary predictions of the model.
In addition to an excellent reproduction of the phase shifts, the potential was able to accurately reproduce deuteron and low-energy 
scattering properties to within 1\%.  
Given the simplicity of this model, we find these results to be very encouraging, 
particularly as regards the quadrupole moment of the deuteron.
The potential is also the first to exhibit the correct high momentum behavior resulting from theoretically motivated form-factors.

A paper which presents the Oxford potential in detail is under preparation by the authors\cite{downum2010qml}.
Nuclear matter calculations have been carried out and show that the Oxford potential has a competitive equation of state.
Future research into the high momentum behavior of the potential is planned.


\begin{theacknowledgments}
Research sponsored in part by the Office of Nuclear Physics, U.S. Department of Energy.
We are grateful for fruitful discussions with Steven Moszkowski, Rupert Machleidt, and Arthur Kerman.
\end{theacknowledgments}



\bibliographystyle{aipproc}   

\bibliography{Bibliography}

\begin{thebibliography}{14}
\expandafter\ifx\csname natexlab\endcsname\relax\def\natexlab#1{#1}\fi
\providecommand{\enquote}[1]{``#1''}
\expandafter\ifx\csname url\endcsname\relax
  \def\url#1{\texttt{#1}}\fi
\expandafter\ifx\csname urlprefix\endcsname\relax\def\urlprefix{URL }\fi
\providecommand{\eprint}[2][]{\url{#2}}

\bibitem[Stoks et~al.(1994)]{stoks1994chq}
V.~G.~J. Stoks, R.~A.~M. Klomp, C.~P.~F. Terheggen, and J.~J. de~Swart,
  \emph{Phys. Rev.} \textbf{C49}, 2950--2962 (1994), \eprint{nucl-th/9406039}.

\bibitem[Wiringa et~al.(1995)]{wiringa1995ann}
R.~B. Wiringa, V.~G.~J. Stoks, and R.~Schiavilla, \emph{Phys. Rev.}
  \textbf{C51}, 38--51 (1995), \eprint{nucl-th/9408016}.

\bibitem[Entem and Machleidt(2003)]{entem2003acd}
D.~R. Entem, and R.~Machleidt, \emph{Phys. Rev.} \textbf{C68}, 041001 (2003),
  \eprint{nucl-th/0304018}.

\bibitem[Machleidt(2001)]{machleidt2000hpc}
R.~Machleidt, \emph{Phys. Rev.} \textbf{C63}, 024001 (2001),
  \eprint{nucl-th/0006014}.

\bibitem[Micu(1969)]{micu1969drm}
L.~Micu, \emph{Nucl. Phys.} \textbf{B10}, 521--526 (1969).

\bibitem[Le~Yaouanc et~al.(1973)]{leyaouanc1973npq}
A.~Le~Yaouanc, L.~Oliver, O.~Pene, and J.~C. Raynal, \emph{Phys. Rev.}
  \textbf{D8}, 2223--2234 (1973).

\bibitem[Liberman(1977)]{liberman1977srp}
D.~A. Liberman, \emph{Phys. Rev.} \textbf{D16}, 1542--1544 (1977).

\bibitem[Shimizu et~al.(2000)]{shimizu2000snn}
K.~Shimizu, S.~Takeuchi, and A.~J. Buchmann, \emph{Prog. Theor. Phys. Suppl.}
  \textbf{137}, 43--82 (2000).

\bibitem[Valcarce et~al.(2005)]{valcarce2005qms}
A.~Valcarce, H.~Garcilazo, F.~Fernandez, and P.~Gonzalez, \emph{Rep. Prog.
  Phys.} \textbf{68}, 965--1041 (2005), \eprint{hep-ph/0502173}.

\bibitem[Downum et~al.(2010)]{downum2010qml}
C.~Downum, J.~Stone, T.~Barnes, E.~Swanson, and I.~Vida\~{n}a, A quark model of
  low energy nucleon-nucleon interactions: I. nucleon-nucleon properties
  (2010), ~To be submitted to Phys. Rev. C.

\bibitem[SAI(2009)]{SAID}
 (2009), the SAID system is a world-wide database of NN and other hadron
  scattering data. Solutions to partial wave analyses maybe be found online at
  \url{http://gwdac.phys.gwu.edu/}.

\bibitem[de~Swart et~al.(1995)]{deswart1995len}
J.~J. de~Swart, C.~P.~F. Terheggen, and V.~G.~J. Stoks, {The Low-Energy
  Neutron-Proton Scattering Parameters and the Deuteron} (1995), unpublished
  Talk, \eprint{nucl-th/9509032}.

\bibitem[Coraggio et~al.(2009)]{coraggio2008smc}
L.~Coraggio, A.~Covello, A.~Gargano, N.~Itaco, and T.~T.~S. Kuo, \emph{Prog.
  Part. Nucl. Phys.} \textbf{62}, 135--182 (2009), \eprint{0809.2144}.

\bibitem[Bogner et~al.(2007)]{bogner2006srg}
S.~K. Bogner, R.~J. Furnstahl, and R.~J. Perry, \emph{Phys. Rev.} \textbf{C75},
  061001 (2007), \eprint{nucl-th/0611045}.

\end{thebibliography}

\IfFileExists{\jobname.bbl}{}
 {\typeout{}
  \typeout{******************************************}
  \typeout{** Please run "bibtex \jobname" to optain}
  \typeout{** the bibliography and then re-run LaTeX}
  \typeout{** twice to fix the references!}
  \typeout{******************************************}
  \typeout{}
 }

 \end{document}